# Los Beneficios del Desarrollo Dirigido por Modelos en los Repositorios Institucionales
## (The Benefits of Model-Driven Development in Institutional Repositories)

Texier, Jose; De Giusti, Marisa; Oviedo, Nestor; Villarreal, Gonzalo L.;Lira, Ariel
jtexier@unet.edu.ve; {marisa.degiusti; nestor; gonzalo; ariel}
@sedici.unlp.edu.ar

Universidad Nacional Experimental del Táchira (UNET), Venezuela.
Servicio de Difusión de la Creación Intelectual, Universidad Nacional de La Plata (SeDiCI), Argentina.

**RESUMEN**
Los Repositorios Institucionales (RI) se han consolidado en las instituciones en las áreas científicas y académicas, así lo demuestran los directorios de repositorios existentes de acceso abierto y en los depósitos diarios de artículos o documentos realizados por diferentes vías, tales como el autoarchivo por parte de los usuarios registrados y las catalogaciones por parte de los bibliotecarios. Los sistemas RI se basan en diversos modelos conceptuales, por lo que en este trabajo se realiza un relevamiento bibliográfico del Desarrollo de Software Dirigido por Modelos (MDD) en los sistemas y aplicaciones para los RI con el propósito de exponer los beneficios de la aplicación del MDD en los RI. El MDD es un paradigma de construcción de software que asigna a los modelos un rol central y activo bajo el cual se derivan modelos que van desde los más abstractos a los concretos, este proceso se realiza a través de transformaciones sucesivas. Este paradigma proporciona un marco de trabajo que permite a los interesados compartir sus puntos de vista y manipular directamente las representaciones de las entidades de este dominio. Por ello, se presentan los beneficios agrupados según los actores que están presentes, a saber, desarrolladores, dueños de negocio y expertos del dominio. En conclusión, estos beneficios ayudan a que todo el entorno del dominio de los RI se concentre en implementaciones de software más formales, generando una consolidación de tales sistemas, donde los principales beneficiarios serán los usuarios finales a través de los múltiples servicios que son y serán ofrecidos por estos sistemas.

**Palabras clave:** repositorios institucionales, desarrollo de software dirigido por modelos, MDD, metamodelos, beneficios.

**ABSTRACT**
The Institutional Repositories (IR) have been consolidated into the institutions in scientific and academic areas, as shown by the directories existing open access repositories and the deposits daily of articles made by different ways, such as by self-archiving of registered users and the cataloging by librarians. IR systems are based on various conceptual models, so in this paper a bibliographic survey Model-Driven Development (MDD) in systems and applications for RI in order to expose the benefits of applying MDD in IR. The MDD is a paradigm for building software that assigns a central role models and active under which derive models ranging from the most abstract to the concrete, this is done through successive transformations. This paradigm provides a framework that allows interested parties to share their views and directly manipulate representations of the entities of this domain. Therefore, the benefits are grouped by actors that are present, namely, developers, business owners and domain experts. In conclusion, these benefits help make more formal software implementations, resulting in a consolidation of such systems, where the main beneficiaries are the end users through the services are offered.

## 1. Introducción

Los repositorios institucionales (RI) están teniendo un gran auge en las comunidades universitarias, principalmente en lo científico y académico. Se agrupan a través de directorios de repositorios existentes de acceso abierto (por ejemplo OpenDOAR o ROAR) y sus usos se evidencian a través de los depósitos diarios de artículos o documentos realizados por varias vías, tales como el autoarchivo por parte de los usuarios registrados y las catalogaciones por parte de los bibliotecarios. El diseño y desarrollo de RI requiere que las diferentes partes interesadas: desarrolladores, dueños del negocio y expertos del dominio, se pongan de acuerdo sobre un lenguaje neutral y de alto nivel para describir, discutir y negociar los servicios que se pueden ofrecer. Por ello, se desea que la construcción de sistemas o aplicaciones para RI se realice bajo metodologías formales con el fin de obtener productos de mayor calidad. Una metodología en crecimiento y soportada por tecnologías abiertas es el Desarrollo de Software Dirigido por Modelos (MDD).

Este trabajo se centrará en exponer los beneficios para los interesados al aplicar MDD en sistemas y aplicaciones para RI. En la actualidad, según OpenDOAR, existen 2234 repositorios en el mundo (para el 21 de agosto del 2012), los cuales se han implementado y desarrollado con base en modelos conceptuales creados en los últimos 25 años (DELOS, FRBR, Norma ISO 14721, entre otros). Por ello, lo ideal para consolidar un sistema de RI es construir un modelo general que disponga de las múltiples virtudes de cada uno de los modelos conceptuales más destacados, los cuales generen un marco de trabajo para aplicaciones relacionadas con este dominio bajo una metodología en particular, que en este caso será MDD. El paradigma MDD permite a los interesados compartir sus puntos de vista y manipular directamente las representaciones de las entidades de este dominio.

Este artículo se estructura de la siguiente manera: primero, una conceptualización de lo que significan los repositorios; segundo, una explicación de la metodología para la construcción de un framework general; tercero, la descripción de los beneficios para el desarrollo de sistemas y aplicaciones de RI bajo la metodología MDD; y, por último, se esbozan unas líneas de trabajos futuros desde el Servicio de Difusión de la Creación Intelectual (SeDiCI) y se detallan las conclusiones del trabajo.

## 2. Los Repositorios Institucionales

Los diferentes conceptos y descripciones de repositorios institucionales han tenido su entrada en el mundo científico desde principios de los años 2000 con autores como Clifford Lynch [1] y Van de Sompel [2]. Los Repositorios Institucionales se entienden como estructuras web interoperables que alojan recursos científicos, académicos y administrativos, tanto físicos como digitales, descriptos por medio de un conjunto de datos específicos (metadatos). Los RI tienen como propósito recopilar, catalogar, gestionar, acceder, difundir y preservar. Para complementar esta definición se pueden enumerar las siguientes características:
- Los RI pertenecen a una institución académica o de investigación.
- Los materiales de las instituciones que representan su producción intelectual deben estar presentes en el RI, dando lugar a una colección de documentos y objetos, de varios tipos y formatos.
- Investigadores afiliados u otras personas pertenecientes a la organización pueden depositar directamente sus textos, conjuntos de datos, archivos de sonido, imágenes o cualquier otro tipo de documento, de manera personal (autoachivo) o a través de los catalogadores.
- Los documentos pueden estar en cualquier etapa del proceso de la investigación académica, esto depende de la política de la institución sobre los documentos que se pretenden depositar.
- Un porcentaje alto de los Repositorios Institucionales están fundamentados en los ideales y objetivos del Open Access.
- Los RI pueden estar agrupados en directorios de repositorios y los directorios más referenciados de acuerdo con los enlaces entrantes o *inlinks*, según Majestic SEO [3] y ahrefs [4], son: OpenDOAR (Directory of Open Access Repositories) [5] con 2234 repositorios registrados, ROAR (Registry of Open Access Repositories) [6] con 2925 repositorios registrados y University of Illinois OAI-PMH Data Provider Registry [7] con 2937 repositorios (datos al 21 de agosto del 2012).

Con frecuencia se confunden las definiciones de los RI con las definiciones de Bibliotecas Digitales (BD) que surgieron en los noventa, con autores como Lesk [8], Waters [9], Borgman [10] y Chowdhury [11]. Es preciso aclarar que en este trabajo no se profundizará en tal distinción, aunque se puede concluir que un repositorio institucional es una biblioteca digital y una biblioteca digital es un repositorio institucional, tal y como lo afirman los autores Xia y Opperman [12]. En cambio en las propuestas de modelos conceptuales para este dominio, no se hace una fuerte distinción si sirven para los RI o BD, ya que el principal objetivo de los modelos es ofrecer una colección de recursos a una comunidad de usuarios mediante el acceso y la transferencia de la información.

Los modelos son un conjunto de elementos que sirven para demostrar la consistencia de una teoría, es decir, representan con detalle un sistema dado. En este trabajo, los modelos de RI están enmarcados por problemas originados por la representación de los recursos y la diversidad de soluciones tecnológicas disponibles en los distintos módulos de los RI, tales como: esquema de metadatos, almacenamiento, arquitectura, catalogación, indexación y preservación de los recursos. Los modelos conceptuales de las BD más destacados en el relevamiento bibliográfico realizado son:
- Marco conceptual propuesto por Bawden y Rowlands. Se basa en una serie de términos esenciales para comprender el concepto de biblioteca digital y sus componentes, realizado en 1999 [13]. Estos autores se centran en tres aspectos cruciales: informacional (documentos), sistemas (tecnología) y social (trabajo).
- Norma ISO:14721, modelo de referencia OAIS (Open Archival Information System). En el año 2003 se publicó la Norma ISO:14721, denominada *ISO Reference Model for an Open Archival Information System* (OAIS) [14], que modela las partes que componen un sistema abierto de archivo de información, por ejemplo un repositorio institucional, sin definir su implementación. Las dos funciones principales del modelo OAIS son conservar la información y garantizar el acceso a la misma.
- El modelo formal propuesto por Goncalves et al., conocido como "Streams, structures, spaces, scenarios, societies" (5S) [15]. Este modelo fue propuesto en el 2004 y se basa en cinco abstracciones fundamentales: Streams, Structures, Spaces, Scenarios and Societies, las cuales ayudan a definir, a relacionar y a unir conceptos de objetos digitales, metadatos, colecciones y servicios, requeridos para formalizar las bibliotecas digitales.
- El modelo general DELOS [16], propuesto por Candela et al. que consiste en un modelo para las bibliotecas digitales a partir del Manifiesto de DELOS dedicado a establecer las

- bases que fundamentan a las bibliotecas digitales a través de la identificación de los conceptos.
- El modelo conceptual FRBR (Functional Requirements for Bibliographic Records) fue propuesto en 1998 por la International Federation of Library Associations (IFLA) [17]. Los FRBR son una representación simplificada del universo bibliográfico, basado en un modelo entidad-relación.
- El metamodelo CRADLE (Cooperative-Relational Approach to Digital Library Environments) [18], desarrollado por Malizia, Bottoni y Levialdi, es un marco basado en un metamodelo y en un lenguaje visual para la definición de conceptos y servicios relacionados con el desarrollo de bibliotecas digitales.

Estos modelos conceptuales permiten sentar las bases para desarrollar un modelo general para los sistemas de Repositorios Institucionales/Bibliotecas Digitales, los cuales pueden desarrollarse bajo un paradigma dirigido por modelos tales como: MDD, DSL, DSM, líneas de productos, etc. En este trabajo se profundiza únicamente sobre la metodología MDD y su relación con los RI.

## 3. Desarrollo de Software Dirigido por Modelos (MDD)

El desarrollo de software dirigido por modelos, conocido en inglés como Model Driven Development (MDD), es un paradigma de construcción de software cuyas motivaciones principales son la independencia de los productores de software a través de estandarizaciones y la portabilidad de los sistemas de software [19]. El objetivo de MDD es separar el diseño del sistema de la arquitectura de las tecnologías, para que puedan ser modificados independientemente. Para lograr esto, se asigna a los modelos un rol central y activo bajo el cual se derivan modelos que van desde los más abstractos a los concretos, este proceso se realiza a través de transformaciones sucesivas e iteraciones. La mayor importancia de este paradigma radica en que todo debe girar sobre la base de modelos, definidos a partir de metamodelos, que ayudan al computador a entenderlos y a transformarlos. El Model Driven Architecture (MDA o Arquitectura Dirigida por Modelos) es una propuesta de MDD definida por Object Management Group (OMG). En ocasiones, el término MDA se intercambia con el de MDD, ya que MDA se refiere a las actividades que llevan a cabo los desarrolladores, mientras que MDD hace referencia a su definición formal [20], por tanto, bajo este contexto, es indistinto hacer referencia a MDA o a MDD.

### 3.1. Ciclo de vida del MDD
El ciclo de vida de desarrollo de software en MDD (ver Figura 1) basa su funcionalidad en tres modelos [21]. Un modelo computacional independiente (Computation Independent Model o CIM) que es una vista del sistema que no muestra detalles de su estructura y se le puede conocer como el modelo del dominio; esto corresponde tradicionalmente a las etapas de captura de requisitos y análisis [19]. Luego, se define un modelo independiente de la plataforma (Platform Independent Model o PIM) a través de un lenguaje específico para el dominio en cuestión e independiente de cualquier tecnología. El modelo PIM puede traducirse a uno o más modelos específicos de la plataforma (Platform Specific Model o PSM). Las implementaciones de los PSM, pueden estar basadas en lenguajes específicos del dominio o lenguajes de propósito general como Java, C#, base de datos relacionales (SQL), Python, etc. Luego de definir cada uno de estos modelos, se realiza la implementación del código fuente (Implementation Model o IM) a partir de cada uno de los PSM desarrollados. Es importante destacar que los PIM o PSM pueden contener varios modelos correspondientes a puntos de vista del sistema.

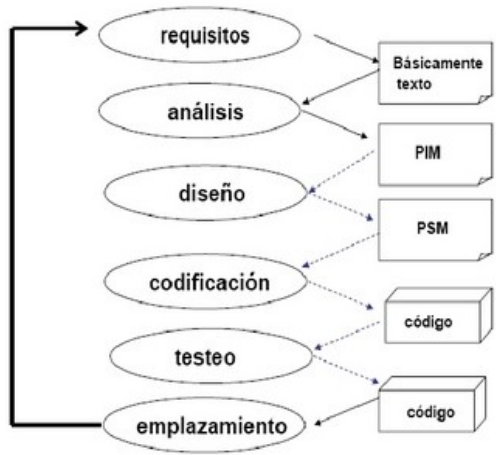

Figura 1. Ciclo de vida de MDD. Fuente: Pons et al., 2010

### 3.2. Arquitectura de Metamodelado de 4 capas en MDA

MDA plantea una metodología basada en la generación de modelos (PIM y PSM), los cuales pueden transformarse en otros modelos sucesivamente hasta obtener una representación final (usualmente código ejecutable) gracias a la definición de un lenguaje, que proporcione la posibilidad de describirlos de forma adecuada. Este lenguaje está a su vez definido a partir de un metalenguaje [20], lo cual genera la idea de sucesivas capas recursivas (ver Figura 2).

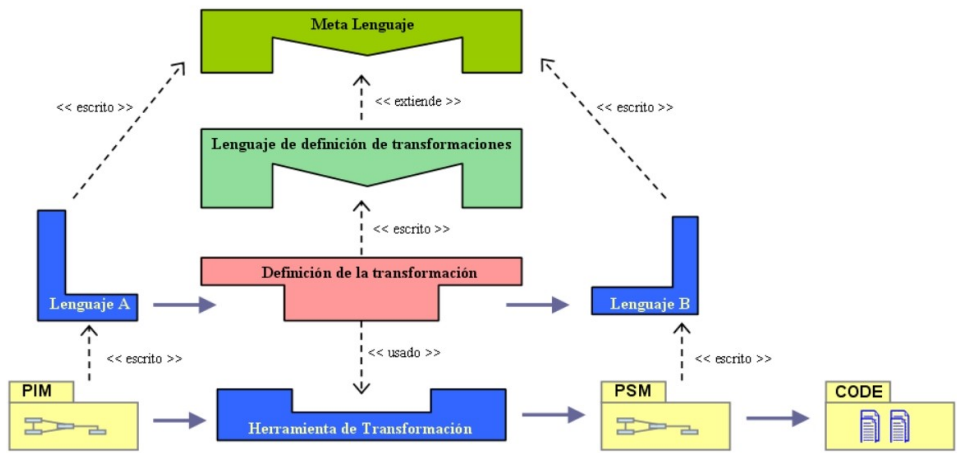

Figura 2. Vista en niveles de la arquitectura de metamodelado. Fuente: Giandini, R., CIbSE 2010

El OMG especifica una serie de capas o niveles de modelado (Figura 3):
- Capa M3: Meta-metamodelo. Define los elementos existentes en la capa M2, mediante instancias de elementos existentes en esta capa. En definitiva, en este nivel está definido el meta-metalenguaje o meta-metamodelo. (*MOF*).
- Capa M2: Metamodelo. Este nivel contiene los elementos del lenguaje de modelado o metamodelo. (*UML/JAVA*).
- Capa M1: El modelo del sistema. A este nivel se define el modelo del sistema o la aplicación propiamente dicha. (*Modelo UML / Modelo Clases Java*).
- Capa M0: Las instancias. Es la capa en donde se ejecuta el sistema, donde están las instancias reales que se han creado durante la ejecución. (*Objeto UML / Instancias de Java*)

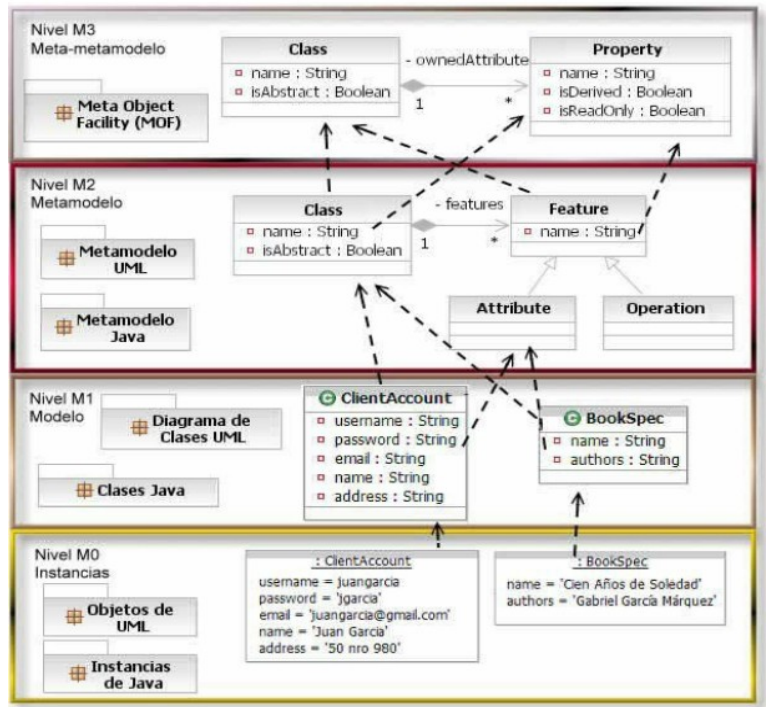

Figura 3. Vista general de las relaciones de las capas. Fuente: Pons et al., 2010

### 3.3. ¿Por qué MDD en los RI?
Los actores involucrados (desarrolladores, dueños del negocio y expertos del dominio) en el diseño y desarrollo de un sistema junto con sus componentes para RI deben ponerse de acuerdo sobre un lenguaje neutral y de alto nivel, que les sirva de apoyo para describir, discutir y negociar los servicios (recopilar, catalogar, gestionar, acceder, difundir y preservar) que el RI debe ofrecer. El paradigma MDD brinda el marco que permite a los interesados compartir sus puntos de vista y manipular directamente las representaciones de las entidades de este dominio. Además, este paradigma ofrece algunas ventajas [19] como: incremento en la productividad (errores, costos, código), adaptación a cambios tecnológicos, reuso de software, mejora en la comunicación con usuarios y desarrolladores, asignación de roles, entre otros.

La utilidad del metamodelo en MDD está centrada en definir lenguajes de modelado sin ambigüedades, en contar con herramientas de transformación para leer y entender los modelos, en tener reglas de transformación claras que describen cómo un modelo en un lenguaje fuente va a ser transformado a un modelo en un lenguaje destino y en el uso de definiciones formales (Figura 2) obtenidas por la sintaxis de los lenguajes, facilitando su automatización.

La aplicación del paradigma MDD al complejo sistema de información de los Repositorios Institucionales permite trasladar las ventajas del MDD y la utilidad del metamodelado al desarrollo de los sistemas de repositorios, que involucran tecnologías y características de áreas como: bibliotecas, sistemas de información, recuperación de información, representación de información e interacción persona-computador [22]. Estos beneficios son capitalizados a su vez en los diferentes procesos presentes en los RI proporcionando a los desarrolladores conceptos del dominio sin ambigüedades.

## 3.4. Experiencias del MDD en otros dominios

En muchas ocasiones, una de las maneras de evaluar el éxito de metodologías de desarrollo de software como la descrita en este trabajo, es observando implementaciones relacionadas en otros dominios. A continuación se presentan algunas experiencias del Desarrollo de Software Dirigido por Modelos que han tenido éxito en otros dominios:

- Almacenes de Datos (Data Warehouse) gracias al metamodelo CWM (Common Warehouse Metamodel), que se utiliza para un intercambio de información de metadatos del almacén de datos [23].
- Modelado de aplicaciones web, por ejemplo el desarrollo de una aplicación de hipermedia móvil [24].
- Integración de Sistemas gestión de aprendizaje, conocidos como LMS (Learning Managment Systems), que permiten principalmente la interoperabilidad de los recursos existentes en al menos dos plataformas [20].
- Desarrollo de aplicaciones usando estándares de metamodelos tales como: Business Process Modeling Notation (BPMN), Ontology Definition Metamodel (ODM) ó Software Process Engineering Metamodel (SPEM) [25].

## 4. Beneficios de aplicar MDD en los RI

Lo ideal para consolidar los sistemas de RI es construir un modelo general que disponga de las diferentes virtudes de cada uno de los modelos existentes más destacados y que generen un marco de trabajo para aplicaciones relacionadas con este dominio bajo una metodología en particular, que en este caso es MDD. El paradigma MDD brinda un marco que permite a los interesados compartir sus puntos de vista y manipular directamente las representaciones de las entidades de este dominio. Por ello, en esta sección se presentan una serie de beneficios para los tres actores presentes en el desarrollo de software que se pueden tomar tanto de la aplicación del paradigma como de su implementación en el contexto de los RI:

*4.1. Los desarrolladores y/o diseñadores:* son los responsables de hacer realidad los requerimientos del sistema a desarrollar, cuyos beneficios son:

- Un menor número de líneas de código escritas, ya que los niveles de abstracción de MDD a través de los modelos y metamodelos diseñados fomentan el reuso del código y de los modelos.
- Alto nivel de abstracción para escribir aplicaciones y artefactos de software a través de la arquitectura de niveles del metamodelado y las capas de modelado de MDA. Este beneficio favorece diseñar una aplicación o artefactos de software partiendo de lo más general a lo más concreto, es decir, son independientes de la tecnología. Por ejemplo, los objetos del repositorio se pueden abstraer a un formato general propio de los RI, además de establecer relaciones entre ellos. Algunos ejemplos de objetos son: documentos, autores, instituciones, tesauros, sistemas de clasificación, áreas temáticas, entre otros.
- Especificación de requisitos de usuario a varios niveles obteniendo un sistema flexible a los cambios. La flexibilidad de estos sistemas se observa si la funcionalidad que se desea agregar es posible implementarla a través de los modelos correspondientes y evidentemente en el código generado.
- Evitar la adopción de una única tecnología de hardware particular gracias a los niveles de abstracción presentes sin generar un vínculo particular, por tanto, los metamodelos se convierten en el eje central, porque representan el modelo del sistema de manera independiente a la tecnología.
- Interoperabilidad entre los objetos en los sistemas de RI en un entorno multiplataforma. Esto se logra por los niveles de abstracción que permiten representar en un modelo general las relaciones entre objetos del RI en diversas plataformas.

*4.2. Dueños del negocio:* son los encargados de coordinar y/o financiar el proyecto de desarrollo e implementación del sistema dentro de la organización o institución, y los beneficios son:
- Desarrollo de componentes de software para los sistemas de repositorios. Este beneficio se evidencia por las fases del MDD, que van del CIM a la generación de código, ayudando a los desarrolladores a crear diseños de artefactos correctos sin causar problemas en el sistema general con el fin de incrementar la productividad y calidad del mismo. Esos artefactos pueden ser indicadores para evaluar el uso del sistema, por ejemplo, recursos descargados y visitados, tiempo de permanencia en vistas de los objetos, etc.
- Preservación digital de los recursos y/o de los objetos generando estrategias que parten de los niveles altos de abstracción, como son los PIM. La importancia de la preservación radica en garantizar la perpetuidad en el tiempo del recurso almacenado.
- Generación de código para plataformas previamente especificadas a través de la funcionalidad que ofrece el paradigma, ya que los artefactos de software se generan en los PSM que luego se transformarán en el código deseado.
- Reducción de costes en el desarrollo de aplicaciones debido a la disminución del recurso humano requerido, de las horas hombre y del tiempo invertido en las diferentes actividades relacionadas.
- Documentación de todo el proceso de desarrollo de software, representado básicamente en el modelo PIM, documentación de alto nivel que se necesita para cualquier sistema de software.

*4.3. Expertos del dominio:* representan a los especialistas presentes en las fases del mundo de los RI. Los beneficios son:
- Permitir el desarrollo de modelos por parte de los distintos expertos del dominio, a diferencia de los desarrolladores que se concentran en los detalles técnicos a través de transformaciones de los modelos partiendo desde los PIM hasta la generación de código.
- Generación de lenguajes específicos del dominio en las fases de la implementación de los RI bajo MDD, tales como: modelo de datos, modelo de la arquitectura, modelo de las entidades abstractas, interfaz de usuario, entre otros. Este beneficio se obtiene en el desarrollo de modelos correspondientes a diferentes puntos de vista para las fases principales del paradigma, PIM y PSM, al mismo tiempo la posibilidad de implementar lenguajes específicos del dominio para esos puntos de vista garantizando el reuso de los conceptos.
- Interoperabilidad entre los distintos modelos PSM principalmente, ya que pueden pertenecer a distintas tecnologías (plataformas). Esta interoperabilidad se logra a través de puentes construidos por las herramientas de transformación de modelos garantizando los conceptos definidos.

Estos beneficios ayudan a que todo el entorno del dominio de los RI se concentre en implementaciones de software más formales, generando una consolidación de tales sistemas, donde los principales beneficiarios serán los usuarios finales a través de los diferentes servicios que son y serán ofrecidos por estos sistemas.

## 5. Conclusiones y Trabajos Futuros

Los beneficios de aplicar el Desarrollo de Software Dirigido por Modelos en sistemas y aplicaciones para RI que se han expuesto en este trabajo se tomaron en cuenta a partir del relevamiento bibliográfico realizado de los modelos conceptuales más importantes y de las experiencias de usar la metodología MDD en el mundo de los RI/BD.

El concepto de los Repositorios ha evolucionado y se ha relacionado con el concepto de una Biblioteca Digital, tal y como se evidencia en sus modelos conceptuales. Además, se destaca el gran auge que han tenido los RI en los últimos años de acuerdo con los directorios de repositorios existentes y con el afianzamiento de la filosofía del acceso abierto en la comunidad de investigadores y académicos. Todas estas realidades incentivan el estudio de este dominio así como también incentivan el desarrollo formal de aplicaciones y componentes de software. De igual manera, se recomienda que todos los trabajos, investigaciones y desarrollos de metamodelos, modelos y artefactos de software, sean desarrollados bajo la filosofía del open source para que exista un crecimiento y una constante retroalimentación de los múltiples productos generados.

El desarrollo formal de desarrollo por modelos es aconsejable realizarlo en el marco de alguna metodología, por ello, este trabajo se centró en el desarrollo de software dirigido por modelos. Se expuso una definición y una breve descripción de la metodología, de igual manera se encontraron trabajos que afirman la consolidación de la misma y de los diversos beneficios de la aplicación de este enfoque y de otros que surgen de la implementación en este contexto.

Evidentemente, se puede seguir ampliando el tema y profundizando en determinados aspectos que se han pasado por alto, por ello, desde SeDiCI se desea dar continuidad al trabajo presentado complementandolo con los siguientes trabajos:
- Analizar, evaluar y comparar las diferentes metodologías dirigidas por modelos que pueden aplicarse para las aplicaciones y/o artefactos de software del dominio tratado en este trabajo.
- Diseñar un metamodelo que permita el diseño de modelos específicos de cada uno de los subprocesos del dominio de los RI/BD bajo el paradigma del desarrollo de software dirigido por modelos propuesto por OMG.

## 6. Referencias Bibliográficas